\documentstyle[epsf]{mn}

\title[Cl of RASS]{Cosmological constraints from the cluster
contribution to the power spectrum of the soft X-ray background. New evidence 
for a low $\sigma _8$}

\author[Diego et al.]
   {J.M. Diego$^1$, W. Sliwa$^2$, J. Silk$^1$, X. Barcons$^3$, W. Voges$^4$ \\
    $^1$ University of Oxford. 
       Denys Wilkinson Building, 1 Keble Road, Oxford OX1 3RH, 
       United Kingdom.\\
    $^2$ Nicolaus Copernicus Astronomical Center, Bartycka 18, 00-716 Warsaw.\\
    $^3$ Instituto de Fisica de Cantabria (CSIC-UC), 39005 Santander, Spain.\\
    $^4$ Max-Planck-Institut f\"{u}r extraterrestrische Physik, 85741 Garching, Germany.
} 
\date{Draft version \today}

\pagerange{\pageref{firstpage}--\pageref{lastpage}}

\begin{document}

\maketitle

\label{firstpage}
\begin{abstract}
We use the X-ray power spectrum of the $ROSAT$ all-sky survey in the
R6 band ($\approx$ 0.9-1.3 keV) to set an upper limit on the galaxy
cluster power spectrum.  The cluster power spectrum is modelled with a
minimum number of robust assumptions regarding the structure of the
clusters.  The power spectrum of $ROSAT$ sets an upper limit on the
$\Omega_m-\sigma_8$ plane which excludes all the models with
$\sigma_8$ above $\sigma_8 = 0.5\Omega_m^{-0.38}$ in a flat
$\Lambda$CDM universe.  We discuss the possible sources of systematic
errors in our conclusions, mainly dominated by the assumed $L_x-T$
relation. Alternatively, this relation could be constrained by using
the X-ray power spectrum, if the cosmological model is known.  Our
conclusions suggest that only models with a low value of $\sigma _8$
($\sigma _8 < 0.8$ for $\Omega _m = 0.3$) may be compatible with our
upper limit. We also find that models predicting lower luminosities in
galaxy clusters are favoured. Reconciling our cosmological constraints
with these arising by other methods might require either a high
entropy floor or wide-spread presence of cooling flows in the
low-redshift clusters.
\end{abstract}


\section{Introduction}\label{section_introduction}
Galaxy clusters are a useful probe of cosmological models.  In recent
years there has been a wide variety of work based on the study of the
population of galaxy clusters as a way to constrain cosmological
models. The mass function derived from N-body simulations as well as
analytical approaches like the Press-Schechter mass function show a
strong dependence of the mass function on the cosmological model. By
using the local abundance of clusters, it is possible to set strong
constraints on the $\sigma_8 - \Omega_m$ plane.  

However, both parameters are degenerate and one needs to go to higher
redshifts in order to break this degeneracy. Combining the low
redshift mass function with the high redshift mass function it is
possible to break the degeneracy in $\sigma_8 - \Omega_m$ and
determine both parameters (Bahcall \& Fan 1998).  Nevertheless, a
measurement of the mass function (especially at high redshift) is very
difficult to achieve since the mass of a cluster cannot be measured
directly but requires modelling of the observations.  Direct
measurement of the mass requires the use of sophisticated techniques
such as gravitational lensing (and high quality data) or tracing the
mass via the X-ray emission under the assumption of hydrostatic
equilibrium.

An alternative approach can be devised by
transforming masses into temperatures or luminosities and using the
equivalent temperature function or luminosity function instead of the
mass function.  The drawback of this last approach is that in this
case a $T-M$ relation or a $L_x-M$ relation needs to be assumed. These
scaling relations suffer from scatter and they are not yet very well
established. Even in the case where these relations are well known,
there are still several systematic effects which are difficult to be
accounted for.  For instance, the luminosity of a cluster depends on
the cosmological model since luminosities are obtained from the
measured flux times the square of the luminosity distance. In the case
of X-rays, the X-ray emission is concentrated in the centre of the
cluster (due to the $n^2$ dependence with $n$ being the electron density) 
and most of the emission coming from the external parts of the cluster is 
not detected. In order to compute the total luminosity of the cluster one 
needs to assume a profile for the electron density and extrapolate the 
observed emission to the outer parts. 
When using the mass, temperature, flux or luminosity
functions in cosmological studies, a precise knowledge of the
selection function of the survey is required since the observed
function will be compared with the theoretical one in which the
integration of the mass (or temperature, flux, luminosity) starts at a
minimum value (as a function of $z$) defined by the selection function
of the survey.  Usually this selection function is inhomogeneous due,
e.g., to different exposures in different sky directions.
This makes the modelling of the theoretical
mass (temperature, flux or luminosity) function very complicated. Also
due to the selection function of the survey, the number of clusters
observed (i.e, detected above some detection threshold) represents only a small
fraction of the population of clusters. This means that cosmological
studies based on the mass (temperature, flux of luminosity) function
are based only on the {\it tip of the iceberg} and therefore subject
to large variations due to small number statistics.\\
\begin{figure*}
   \begin{center}
   \epsfysize=11.cm 
   \caption{
            ROSAT all-sky survey (RASS) map in the R6 band in galactic coordinates. 
            The map has been 
            repixelized using \small{HEALPIX}\normalsize. The colour scale is logarithmic 
            and the units are $10^{-6} cts/s \ arcmin^2$. 
            The two horizontal lines mark the two constant Galactic latitudes at 
            $|b| = 40^{\circ}$. The thick lines show the galactic 
            meridians $l=70^{\circ}$ (left) and $l=250^{\circ}$ (right). 
           }
   \label{fig_RASS}
   \end{center}
\end{figure*}
In this work we propose the use of the power spectrum as an alternative 
cosmological test which is less affected by the problems pointed out above. 
Our attempt is not to measure the power spectrum of galaxy clusters (see for 
instance Schuecker et al. 2001) but to use the power spectrum of 
the diffuse X-ray background as an upper limit 
(see Fabian \& Barcons 1992 for a discussion of the X-ray background). 
This approach has several advantages over previous work. \\
(i) First, since we use the diffuse power spectrum as an upper limit, 
we do not need to {\it detect} the clusters. The only thing we need to 
do is to model the X-ray cluster power spectrum and compare it with the measured 
one. The cluster power spectrum must remain below the one due to the global 
fluctuations of the diffuse cosmic X-ray background (Barcons \& Fabian 1988, 
Carrera et al. 1997, Barcons et al. 1998, Yamamoto \& Sugiyama 1998).\\
(ii) In modelling the cluster power spectrum we do not need to know the selection 
function since all the clusters (detected and undetected) 
will contribute to the power spectrum. \\
(iii) The previous point has a very interesting feature. 
We are not only sensitive to the tip of the iceberg but to the whole iceberg 
itself! This fact has interesting applications in studies like pre-heating 
(see e.g Voit et al. 2002) which are particularly important in the low-mass 
regime where clusters are difficult to detect. \\
(iv) Since we do not need to detect the individual clusters 
we are not affected by any extrapolation of the measured flux (needed to get 
the total luminosity or mass) making our results more independent of the assumed 
profile of the clusters. \\
(v) The power spectrum is scale dependent. Previous work has used the energy 
spectrum as an upper limit to constrain the emission from galaxy clusters 
(see e.g Wu et al. 2001). However, in these studies the measured emission is 
contaminated by the diffuse emission coming from the Galaxy. This is equivalent 
to studying just the monopole of the multipole decomposition.
On the contrary, the power spectrum can disentangle the diffuse from the point-source
emission since the diffuse emission will contribute significantly only to the 
large scales (low multipoles) while the emission from point-sources
will dominate the
smaller scales (intermediate-high multipoles). \\

In the next sections we will describe briefly the power spectrum of RASS diffuse 
emission (section \ref{sect_ROSAT_Cl}), 
we will define the model to compute the cluster power 
spectrum (section \ref{sect_theor_Cl}) and we will constrain the cosmological model 
(section \ref{sect_constraints}). In section \ref{sect_systematic} we revise the 
assumptions made in the previous sections and check how the results change 
when we change the model. Finally, we  discuss our results in 
section \ref{sect_discussion} and summarise in section \ref{sect_conclusion}.
Throughout the paper (unless otherwise stated), 
the Hubble constant is given as 100 $h$ km s$^{-1}$ Mpc.  
\section{The power spectrum of ROSAT}\label{sect_ROSAT_Cl}
The use of the spherical harmonics decomposition of the intensity of
the X-ray background as a means to measure or constrain cosmological
parameters was first proposed by Lahav, Piran \& Treyer (1997). The
emphasis was mostly on the contribution of the clustering of point
sources (mostly AGNs). They focused their analysis on the low 
multipoles ($l < 10$).  
That sort of study has the disadvantage that
the physics that leads to the formation and clustering of these
exceptional sources is not simple and it is hidden in an unknown {\it
bias} parameter. Indeed, Treyer et al. (1998) conducted that analysis
on the 2-10 keV data from the HEAO-1 A2 experiment and found some
interesting constraints on the bias parameter of X-ray sources, but no
direct implications on the cosmological parameters themselves. The
use of a softer band (but not heavily contaminated by the Galaxy) to
constrain the cluster power spectrum and derive cosmological
parameters is a much more promising task, as the physics of the
cluster formation is much simpler and closer to the Cosmological
model.\\

The harmonic power spectrum of the diffuse $ROSAT$ All-Sky Survey 
(RASS, see Snowden et al. 1997) X-ray background 
has been computed up to $l\approx 1000$ in a previous work (Sliwa et al. 2001).
In that work the authors computed the power spectrum in different areas 
(subsamples) of the sky centred in the north pole. They also studied 
the effects of the data binning in pixels and calculated the error estimates 
due to limited counting statistics, instrumental background subtraction, and cosmic
variance. \\

On the other hand, the power spectrum of clusters in X-rays has been recently measured 
using $ROSAT$ data (see Schuecker et al. 2001 for an estimation based on the REFLEX 
survey and Borgani et al. 1999 for an estimation of its Fourier transform, 
the two point correlation function, from the XBACs cluster sample). 
That measured cluster power spectrum has been also used to constrain the cosmological 
model (Schuecker et al. 2002). In that work, the modelling of the observed cluster 
power spectrum (above certain limiting flux) is a complicated procedure which involves 
several assumptions (such as  the $L_x-M$ relation which was taken from 
Reiprich \& B\"{o}hringer 2002). 
Results obtained in a different study (Viana et al. 2002) seem to indicate 
that there could be an inconsistency between the REFLEX luminosity function and the $L_x-M$ 
relation obtained by Reiprich \& B\"{o}hringer (2002). \\
In this work, instead of detecting clusters and finding the power spectrum from the 
sample of detected clusters, we will use the RASS power spectrum to set an upper limit 
on the cosmological model. This approach has several advantages outlined in the previous 
section. In brief, we recall that the main advantage of this work will be a 
better control of the systematic errors.\\
\begin{figure}
   \begin{center}
   \epsfysize=6.cm 
   \begin{minipage}{\epsfysize}\epsffile{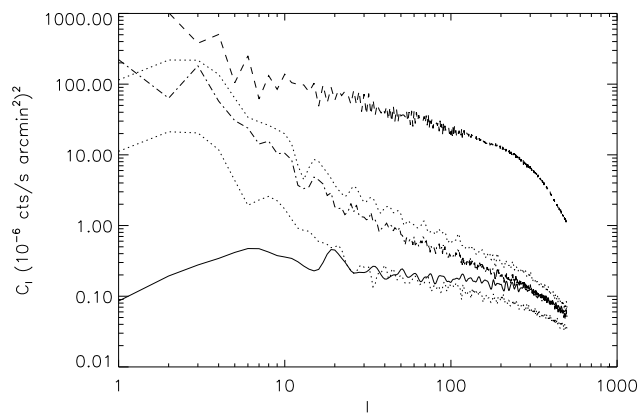}
   \end{minipage}
   \caption{
           $ROSAT$ power spectrum in the R6 band computed 
           with \small{HEALPIX}\normalsize\  in different areas of the sky. 
           Top dashed line $C_l$, for the all-sky RASS data. Dotted lines, 
           $C_l$ for the fraction of the sky with $b > 40^{\circ}$ (top) and  
           $b < -40^{\circ}$ (bottom). Dot-dashed line, $C_l$ 
           for the part of the sky defined by $|b| > 40^{\circ}$ and thick solid line, 
           $C_l$ for the optimal area: $l \in [70^{\circ},250^{\circ}]$ 
           and $b > 40^{\circ}$ (equal to 
           sample D in Sliwa et al. 2001). The power in this area does not 
           show a significant galactic contamination.
           Our estimations are comparable 
           to the ones obtained in Sliwa et al. (2001).
           }
   \label{fig_ROSAT_Cl}
   \end{center}
\end{figure}
We have re-estimated the power spectrum in an independent way (compared with the estimation of 
Sliwa et al. 2001). In our process, we have repixelized the RASS data using the package 
\small{HEALPIX}\normalsize\footnote{available at http://www.eso.org/science/healpix. 
Copyright 1997 by Eric Hivon and Krzysztof M. Gorski. All rights reserved.}\  
and computed the power spectrum with the same package. 
\small{HEALPIX}\normalsize\  was originally designed to analyse mega-pixel CMB data and 
is optimised for the computation of the harmonic power spectrum.\\
The pixel size of the RASS maps is 12 arcmin and the maps are given 
in different bands.
We choose the band R6 which is the one in which the cluster 
signal is expected to be at its maximum compared to the other components 
(local bubble, galactic diffuse emission, extragalactic AGN's). This band is also 
the best in terms of the instrumental response and background contamination.  
We repixelise the RASS in \small{HEALPIX}\normalsize\  
using a pixel size of $\approx 13.7$ arcmin (Nside = 256) 
and compute the power spectrum with the \small{HEALPIX}\normalsize\
subroutine anafast. Nside = 256 gives the closest \small{HEALPIX}\normalsize\  
pixelisation to the original ROSAT's one. 
Due to the large pixel size, we can compute the power spectrum up 
to only $l = 500$. Beyond that point we do not have information about smaller scales.
With this pixel size, the effect of the PSF of $ROSAT$ becomes negligible and we 
will not include it in our analysis. However, at smaller scales, the $ROSAT$ PSF can 
have an important effect on the measured power spectrum.\\
In figure \ref{fig_RASS} we show the RASS data in the R6 band after repixelisation with 
\small{HEALPIX}\normalsize. Several features can be observed in the map,
the most prominent one being the Galactic plane.
The diffuse X-ray maps used in this work have been cleaned of the most prominent point 
sources (see Snowden et al. 1997).
The brightest point sources (Voges et al. 1999, Voges et al. 2000) 
were removed making use of the full resolution of the 
PSPC detector ($1.6'\times 1.6'$ pixel size) using a minimum source-excision radius of 
$3.5'$. 50000 sources where removed above a threshold of 0.02 cts/s in the R6+R7 band.
Below this threshold, there are still many point sources  
which could contribute to the power spectrum. 
In the maps, there are are several strips with no data (black bands in figure \ref{fig_RASS}). 
There are also many pixels with 0 count-rate, specially in the southern hemisphere.\\

We are interested in computing an estimate of the power spectrum in an
area of the sky which is large enough to avoid cosmic variance
fluctuations but also that has a low foreground contamination and low
noise level. The contamination level is determined basically by the
amount of Galactic emission. We should then avoid regions with strong
galactic emission.  The noise level is determined by the exposure
time. We should then concentrate on those regions having low galactic
emission and large integration times. We have selected a region with
$b > 40^{\circ}$ and $l \in [70^{\circ},250^{\circ}]$ (totalling
$\approx 2400$ deg$^2$) as the optimal region for our analysis (see
figure \ref{fig_RASS}). Several groups have selected also this region for 
their studies of the X-ray background (Soltan et al. 1996, Miyaji et al. 1996, 
Kneissl et al. 1996, Sliwa et al. 2001).\\
 
In figure \ref{fig_ROSAT_Cl} we show the power spectrum in different
areas of the sky.  We have represented the power spectrum in different
situations where we exclude (or include) the galactic plane ($|b|$
above or below $40^{\circ}$). The galaxy contributes on large scales
(low $l$'s) but also introduces power at intermediate $l$'s. The power
spectrum calculated in our optimal area is shown as a thick solid
line. In this case, no structure due to the Galaxy is seen, suggesting
that this area may be dominated by the cosmological signal (galaxy
clusters and faint AGNs). The optimal area has also a very low $N_H$
column density, typically $\sim 2\times 10^{20}\, {\rm cm}^{-2}$.  The
absorption of the X-ray flux by this column density is only a few
percent in most of the pixels in this area, an effect that can be
safely neglected.

\section{The cluster power spectrum}\label{sect_theor_Cl}
We have used the halo model to compute the power spectrum due to galaxy 
clusters (Cole \& Kaiser 1988, Bartlett \& Silk 1994, 
Komatsu \& Kitayama 1999, White 2001, Cooray 2002, Komatsu \& Seljak 2002). 
In this model, the power spectrum is due to two main contributions, 
the two-halo contribution and the single-halo contribution. The first 
one includes the correlation between clusters and only contributes significantly in  
the very large scale regime.  
The single-halo contribution accounts for the individual contribution of each cluster 
to the power spectrum and dominates at smaller scales. \\
The large scales (low $l$) of the power spectrum in the optimal area may be affected by 
window effects and cosmic variance (our optimal area is just $\approx 9 \%$ of the sky). 
Although the power spectrum in the optimal area seems to be free of galactic contamination, 
it may happen that the galaxy is still contributing in this area. 
For all these reasons, the low $l$ regime should be excluded in our cosmological 
analysis. We will concentrate only on the intermediate-small scales for which one can 
consider only the single-halo contribution and neglect the two-halo contribution.\\
However, we should keep in mind that, as noted in Komatsu \& Kitayama (1999), 
the clustering contribution can amount to $20\%-30\%$ of the single halo-contribution 
at degree angular scales ($l\approx 100$) (see also Lahav et al. 1997 for smaller 
multipoles).  
Therefore, by neglecting the two-halo contribution we are being conservative in our assumptions. 
The real power spectrum of galaxy clusters will be about $20\%-30\%$  larger than the one 
considered in this work. \\

The single-halo contribution is just an integrated effect over redshift of the individual 
contribution from each cluster.
\begin{equation}
C_l = \int dz \frac{dV(z)}{dz} \int dM \frac{dN(M,z)}{dM} p_l(M,z)
\label{eqn_Cl_cluster}
\end{equation}
where $dV(z)/dz$ is the volume element, $dN(M,z)/dM$ is the cluster mass function 
and $p_l(M,z)$ is the power spectrum (multipole decomposition) of the 
X-ray 2D profile of the cluster with mass $M$ at redshift $z$. 
In this work we have assumed the Press-Schechter approach for the mass function 
(Press \& Schechter 1974) and we use a numerical fit to the multipole decomposition 
of the X-ray 2D profile as we will describe in the next subsection. 
\subsection{The multipole decomposition of the X-ray 2D profile}
\begin{figure}
   \begin{center}
   \epsfysize=6.cm 
   \begin{minipage}{\epsfysize}\epsffile{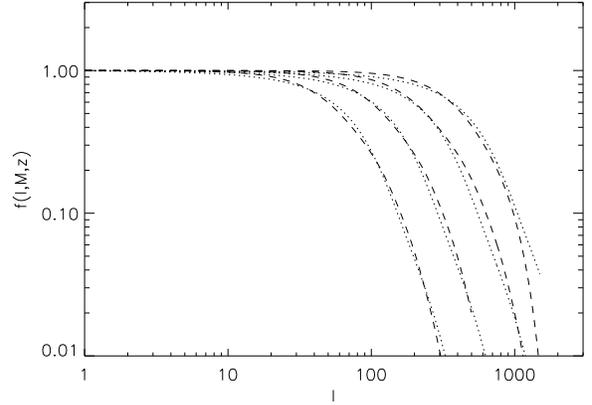}\end{minipage}
   \caption{
            Multipole decomposition of the 2D X-ray profile for different 
            core radii ($r_c$). From left to right. $r_c = 27,13.7,7,3.5$ 
            arcmin. 
            The dashed line represents the real multipole decomposition and 
            the dotted line is the numerical fit of equation \ref{eqn_fit_Cl}. 
           }
   \label{fig_fit_fl}
   \end{center}
\end{figure}
Since the power spectrum of galaxy clusters is proportional to the
square of the density profile, it is important to be careful in the
selection of this profile and make the {\it safest} assumptions about
it.  The usual approach is to consider a numerical model for the
electron density profile (a $\beta$-model or a numerical fit to N-body
simulations) and compute the flux from the integrated signal of the
profile. Then, assume isothermality of the plasma and some relation
between the mass of the cluster and its temperature.  In this process
there are several {\it dangerous} steps.  Assuming a bremsstrahlung
spectrum, the monochromatic volume emissivity of the intra-cluster
medium scales as $j_{\nu} \propto n^2/\sqrt{T}$, where $n$ is the
electron density and $T$ its temperature. Then the integrated
bolometric luminosity scales as the square of the central density
times the square root of the temperature (assuming the cluster is
isothermal). The central density has been poorly estimated so far in
galaxy clusters and there is a lot of uncertainty in its typical
value.  If the cluster is assumed to be isothermal and the scaling
relation between mass and temperature is taken to be the one predicted
by the spherical collapse model, one ends up with a $L_x-T$ relation
which is inconsistent with recent observations. We can conclude that,
modelling of the X-ray emission in galaxy clusters from numerical
models can be a quite unsafe process. It is more useful to formulate
the model in terms of observational quantities rather than theoretical
models which fail in explaining the observations.\\ Basically what we
need for our model is just an expression of the form;
\begin{equation}
p_l(M,z) = p_o(M,z)*f(l,M,z)
\label{eqn_pl}
\end{equation}
where $p_o$ is the normalisation and $f(l,M,z)$ contains the angular dependence 
of the multipole decomposition. To understand why this simple form
will suffice, it is useful 
to think in terms of  a single point source. This source will have a constant 
power spectrum ($f(l,M,z) = 1$) and a normalisation which is equal to;
\begin{equation}
p_o(M,z) = 4\pi|Mean|^2 =4\pi\left | \frac{S_T}{4\pi} \right |^2 
\label{eqn_po}
\end{equation}
where $Mean$ is the total signal, $S_T$, divided by the area of the sky. 
In the case of a cluster, the situation is similar except that in this case 
the source will be extended and $f(l,M,z)$ will not be constant anymore. 
In this case, we have to assume a profile for the electron gas.
We assumed a $\beta$-model (with $\beta = 2/3$) profile 
for the electron cluster density. This assumption is supported by a
wealth of observational data. We also have truncated the profile at the virial radius 
defined as $r_v = p r_c$ with $p$ a fixed parameter ($p \approx 10$) and $r_c$ the 
core radius. 
\begin{equation}
n(r) = \frac{n_o}{1 + (r/r_c)^2}
\end{equation}
Instead of fixing the central density, $n_o$, 
we will fix the total luminosity so the value of $n_o$ will be 
irrelevant in our model. The only free parameter is the core radius, $r_c$.
This will be an advantage compared to other works  
since we can make our model consistent with real observations rather 
than with simulations.\\
To compile an expression for the multipole dependence of the 2D profile, 
we have computed $f(l,M,z)$ for different clusters changing the only free 
parameter, $r_c$, and then fitting the result to an analytical form. 
The best fit we found is;
\begin{equation}
f(l,M,z) = \frac{1}{2}\left(\exp(-\xi_{l,r_c}) + \exp(-\sqrt{\xi_{l,r_c}}) \right)
\label{eqn_fit_Cl}
\end{equation}
with, 
\begin{equation}
\xi_{l,r_c} = l^2 r_c^{1.5/(0.815 + 0.35r_c)}
\end{equation}
where the core radius, $r_c$, is given in rads (see figure \ref{fig_fit_fl}). \\
The mass and redshift dependence of $f(l,M,z)$ is in $r_c$. 
\begin{equation}
r_c = \frac{r_o}{p} M^{1/3} (1 + z)
\label{eqn_rc}
\end{equation}
Although we have assumed the expected redshift dependence of the 
self-similar spherical collapse model with $\Omega_m = 1$, we will adopt this 
form for simplicity. As we will show below, the power spectrum of galaxy 
clusters will be dominated by the low redshift population for which the 
specific form of the redshift dependence is not very important. 
The value of $r_o$ must be assumed and we should keep in mind that typical 
core radii for massive clusters are around 100 kpc. We will adopt a value of 
$r_o/p = 130\, h^{-1}$ kpc for a mass of $M = 10^{15} h^{-1} M_{\odot}$ at redshift 0 
and will discuss other values later.\\
Regarding the normalisation of the profile, $p_o$, we only have to determine 
$Mean$, that is, the total signal divided by the area of the sky. 
To avoid systematic effects due to unrealistic modelling based on simulations, 
we have parametrised the total signal in terms of observed quantities, 
like the $L_x-T$ relation. 
We used the scaling relations derived in Diego et al. (2001). In that work, the 
authors made a consistent fit to the mass function, temperature function, 
X-ray flux function and X-ray luminosity function. In the fit both, the 
cosmological model, and the cluster scaling relations were considered as free 
parameters and were fitted to the data. Using the $L_x-T$ relation we can obtain 
the total signal (flux) of the cluster, $S_T$, from its temperature.
\begin{equation}
S_T = f_c \frac{L_x}{4\pi D_l(z)^2} = 
f_c \frac{L_o T_{keV}^{\alpha} (1 + z)^{\psi}}{4\pi D_l(z)^2}
\label{eqn_ST}
\end{equation}
\begin{figure}
   \begin{center}
   \epsfysize=6.cm 
   \begin{minipage}{\epsfysize}\epsffile{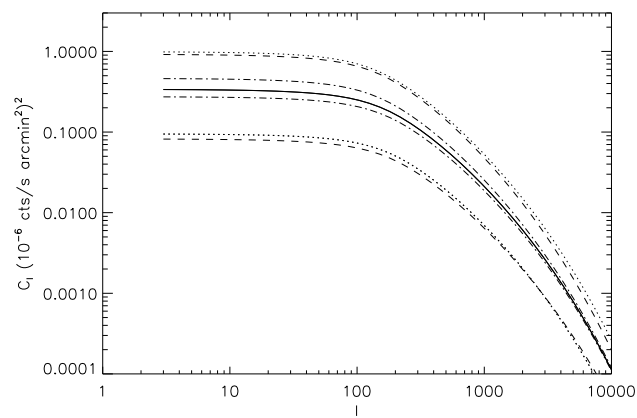}
   \end{minipage}
   \caption{
            Dependence of the X-ray power spectrum on the cosmological parameters 
            $\sigma _8$, $\Omega$ and $\Gamma$.
            The solid line is the reference model ($\sigma _8 = 0.8$, $\Omega _m = 0.3$ 
            and $\Gamma = 0.2$). 
            We show the effect of changing one of the three parameters while keeping 
            unchanged the other two. The effect of changing $\sigma _8$ while 
            $\Omega$ and $\Gamma$ are fixed to the values of the reference model 
            is shown in the dotted lines.
            Bottom dotted line is for $\sigma _8 = 0.7$ and top dotted line is for 
            $\sigma _8 = 0.9$. 
            Dashed lines show the effect of changing $\Omega _m$ in 0.1 units. 
            Bottom dashed line is for $\Omega _m = 0.2$ while the upper dashed line 
            is for $\Omega _m = 0.4$. Finally, the dot-dashed lines show the effect of 
            changing the shape parameter, $\Gamma$. Bottom dot-dashed line is for 
            $\Gamma = 0.25$ and upper dot-dashed line for $\Gamma = 0.15$
           }
   \label{fig_Cl_dependence_S8OmGm}
   \end{center}
\end{figure}
where we use the values of $L_o$, $\alpha$ and $\psi$ obtained in Diego et al. (2001) 
($L_o = 1.12\times 10^{42} h^{-2} erg/s$, $\alpha = 3.2$ and $\psi = 1$ for a 
flat $\Lambda$CDM universe). The temperature is obtained using the $T-M$ relation found in 
Diego et al. (2001) ($T_{keV} = 9.48 M_{15}^{0.75}(1+z)$). This relation is different 
from the best fitting function to the observed $T-M$ relation (see e.g in 
Nevalainen et al. 2000). However, we will use this relation to be consistent with the 
above $L_x-T$ relation and the constraints in Diego et al. (2001). In section
\ref{sect_beyond} we will discuss other (possibly  more realistic) alternatives.\\
We should note that by changing the parameters $L_o$, $\alpha$ and $\psi$ 
we can include also the uncertainty in the $T-M$ relation into an uncertainty in the $L_x-T$ relation. 
That is, the parameters $L_o$, $\alpha$ and $\psi$ will somehow mimic
variations in  the 
parameters of the $T-M$ relation. This choice  is not arbitrary. We could easily model 
the total luminosity as a function of mass and use the $L_x-M$ relation 
as our {\it free-parameter} scaling relation. However, we prefer to parametrise our 
model in terms of the $L_x-T$ relation since it is observationally better constrained 
than the $L_x-M$ relation.\\
The only free parameters in our model are then the $L_x-T$ relation and the 
$r_c$ modelling. We will study their effect later.\\
The factor $f_c$ is the conversion factor between ${\rm erg}\, {\rm
cm}^{-2}\, {\rm s}^{-1}$ to 
$10^{-6} {\rm cts/s}\,  {\rm arcmin}^2$ which are the units in which the RASS data are given. 
We compute this factor with the X-ray package \small{XSPEC}\normalsize\  (Arnaud 1996). 
We have assumed a {\it z-bremsstrahlung} model and considered only the $ROSAT$ 
PSPC channels of the R6 band (channels 91-131, see Snowden et al. 1997). 
The $ROSAT$ response was modelled using the response 
matrix file of the detector PSPC-C. Using a Raymond-Smith for the model  
spectrum with 0.3 solar metallicity in the R6 band 
does not introduce a significant difference compared to the z-bremsstrahlung. 
The $K$-correction was also included in the factor $f_c$. 
This factor was computed for each temperature and redshift. \\ 
As an example, \small{XSPEC}\normalsize\  predicts a conversion of 
1 $cts/s = 8.6\times 10^{-12} erg/s\ cm^{-2}$ in the R6 band (for $kT = 7$ keV and $z=0.5$) \\

\subsection{Cosmological dependence of $C_l$}
The power spectrum shows an important dependence with the cosmological parameters  
$\sigma _8$, $\Omega$ and $\Gamma$. 
In figure \ref{fig_Cl_dependence_S8OmGm} we show some examples where we vary the 
normalisation of the power spectrum, $\sigma _8$, the matter density, $\Omega$, 
and the shape parameter, $\Gamma$. 
The dependence with $\sigma _8$ and $\Omega$ is very important. A change of 
0.2 units in $\Omega$ leads to a change in the power of one order of magnitude. 
Something similar happens with $\sigma _8$ although in this case the change 
is slightly weaker. The dependence on $\Gamma$ is not so important. In particular, 
in our region of interest ($l \approx 100$), $\Gamma$ does not have a very significant effect 
on the power spectrum. For this reason we will fix  $\Gamma$ in our analysis to its 
most favoured value $\Gamma=0.2$ (from 2dF, Percival et al. 2001) and we will vary only 
$\sigma _8$ and $\Omega _m$.
\subsection{Mass and redshift dependence of $C_l$}
Before comparing the cluster power spectrum with the $ROSAT$ data in the R6 band, 
it is interesting to look in more detail at the way the different cluster 
populations contribute to the power spectrum. Given a cosmological model, 
the cluster mass function is just a function of two variables, mass and redshift. 
We can now examine the cluster population at different mass and
redshift intervals to find out which regimes dominate at different scales. \\
\begin{figure}
   \begin{center}
   \epsfysize=6.cm 
   \begin{minipage}{\epsfysize}\epsffile{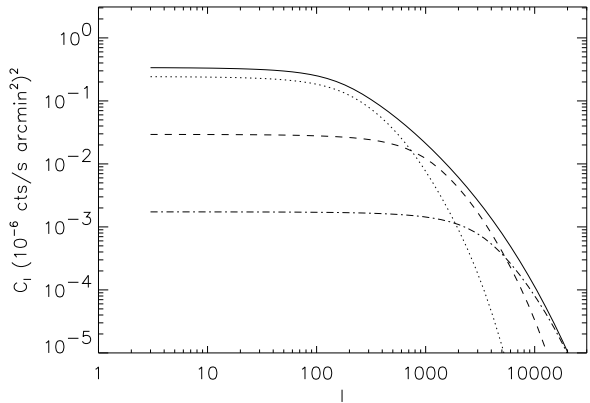}
   \end{minipage}
   \caption{
           Power spectrum of galaxy clusters 
          (with $M \in [3\times 10^{13},10^{16}] h^{-1} M_{\odot}$) 
           split in different redshift intervals  
          ($\sigma _8 = 0.8$, $\Omega _m = 0.3$).
           Solid line $z \in [0.01,2]$, dotted line 
           $z \in [0.01,0.05]$, dashed line $z \in [0.05,0.2]$ and 
           dot-dashed line $z \in [0.2,2]$. 
           }
   \label{fig_Cl_z}
   \end{center}
\end{figure}
\begin{figure}
   \begin{center}
   \epsfysize=6.cm 
   \begin{minipage}{\epsfysize}\epsffile{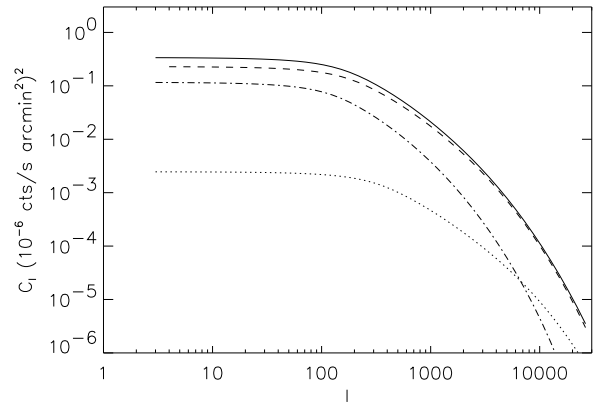}
   \end{minipage}
   \caption{
            Power spectrum of galaxy clusters (with $z \in [0.01,2]$) 
            split in different mass intervals 
            ($\sigma _8 = 0.8$, $\Omega _m = 0.3$).
            Solid line $M \in [3\times 10^{13},10^{16}] h^{-1} M_{\odot}$. 
            Dotted line shows the contribution of the {\it dark clusters}, 
            $M \in [3\times 10^{13},10^{14}] h^{-1} M_{\odot}$.  
            Dashed line shows the dominant contribution of the 
            intermediate-mass clusters, $M \in [10^{14},10^{15}] h^{-1} M_{\odot}$ 
            and the dot-dashed line shows the contribution of the 
            very massive clusters, $M \in [10^{15},10^{16}] h^{-1} M_{\odot}$.
           }
   \label{fig_Cl_mass}
   \end{center}
\end{figure}
Low redshift clusters will dominate the large scales since at low 
redshift the clusters appear larger (figure \ref{fig_Cl_z}). 
Also, it is difficult to find large clusters 
at high redshift. They usually appear at low $z$.
On the contrary, high redshift clusters will dominate the high-$l$ regime. At high $z$, 
clusters will appear with small angular scales. Also, the typical mass of a cluster decreases 
with redshift. However, clusters beyond $z\approx 0.2$ will contribute only at the very small 
scales ($l > 2000$ or scales smaller than $\approx 5\, {\rm arcmin}$). At scales larger than 5 arcmin, 
only the clusters below $z \approx 0.2$ are relevant in the power spectrum.
We also show the contribution in the range $z \in [0.01,0.05]$. 
Many Abell clusters will contribute to this part of the power spectrum 
(min $z_{Abell} \approx 0.01$). 

In terms of the masses (figure \ref{fig_Cl_mass}), the power spectrum is dominated by 
intermediate mass clusters (dashed line). The very massive clusters are rare and 
they do not dominate the power spectrum at any scale. They are big clusters 
and usually appear at low redshift. As a consequence, they will contribute significantly  
only to the large scales.
On the contrary, intermediate-mass clusters are common in a wider range 
of redshifts and their integrated signal dominates the signal of the very 
massive clusters. The intermediate-mass clusters are bright enough to dominate the 
power spectrum at practically all scales. 
Finally, the low-mass clusters will contribute only to the very small scales 
(high $l$-s). They are more abundant than the intermediate-mass clusters but the 
signal is weaker. 

\begin{figure}
   \begin{center}
   \epsfysize=6.cm 
   \begin{minipage}{\epsfysize}\epsffile{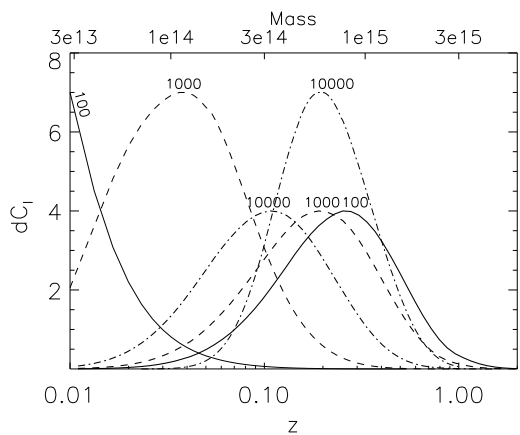}
   \end{minipage}
   \caption{Differential contribution to different multipoles as a 
            function of redshift (top curves) and mass (thick bottom curves). 
            Solid line is the contribution to the multipole $l=100$, 
            dashed line for $l=1000$ and dot-dashed for 
            $l=10000$. All the curves have been normalised to 7 (top) and 
            4 (bottom). The numbers on top of the curves indicate the multipole 
            to which they are contributing. The mass scale is shown in the upper 
            x-axis and the $z$-scale in the bottom x-axis. 
           }
   \label{fig_dCl}
   \end{center}
\end{figure}

Combining the results of figures \ref{fig_Cl_z} and \ref{fig_Cl_mass} we can conclude that 
the cluster power spectrum up to $l \approx 500$ will be dominated by the cluster population 
with $z < 0.2$ and intermediate masses, $M \in [10^{14},10^{15}] h^{-1} M_{\odot}$. 
This can be seen more clearly in figure \ref{fig_dCl} where we present the 
differential contribution to different multipoles as a function of mass (lower set of 
curves) and redshift (upper set of curves). For representation purposes, all the 
power spectra has been normalised to 7 for the redshift case and to 4 for the mass case.

\section{Constraints on $\sigma_8 - \Omega_m$}\label{sect_constraints}
\begin{figure}
   \begin{center}
   \epsfysize=8.cm 
   \begin{minipage}{\epsfysize}\epsffile{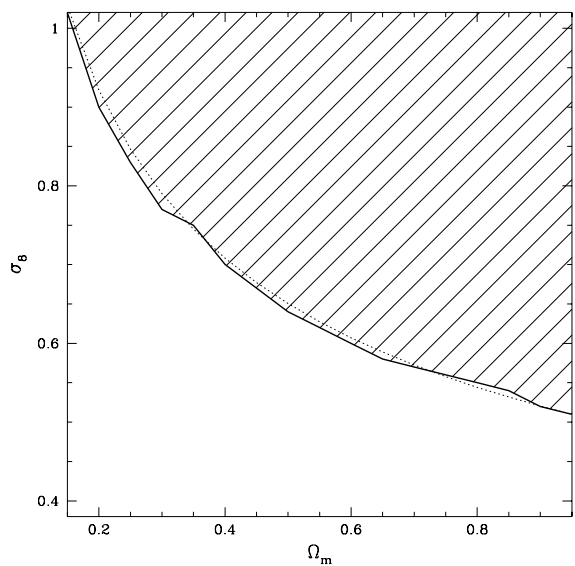}\end{minipage}
   \caption{
            Upper limit constraints in the $\sigma _8 - \Omega _m$ plane (solid line). 
            The dotted line corresponds to the curve, $\sigma_8 = 0.5\Omega_m^{-0.38}$. 
           }
   \label{fig_Om_S8}
   \end{center}
\end{figure}

\begin{figure}
   \begin{center}
   \epsfysize=6.cm 
   \begin{minipage}{\epsfysize}\epsffile{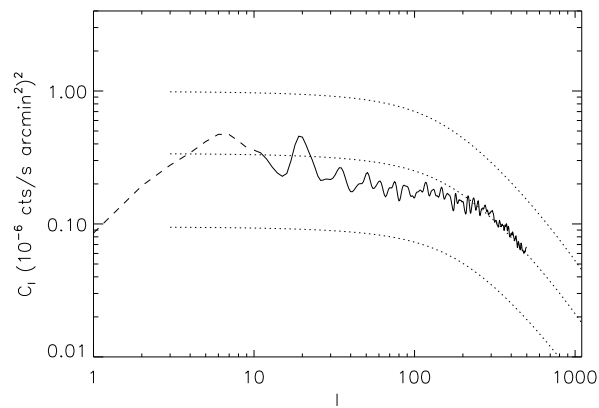}\end{minipage}
   \caption{
           $ROSAT$ power spectrum in the R6 band (solid line) compared with 
           three models above and below the upper limit (dotted lines). 
           From top to bottom, $\sigma _8=0.7$, $\sigma _8=0.8$ and $\sigma _8=0.9$. 
           All models are for $\Omega _m = 0.3$. 
           The dashed line part of the $ROSAT$ power 
           spectrum was not used in the calculation of the upper limit in 
           figure \ref{fig_Om_S8}.
           }
   \label{fig_ROSAT_Cl_S808_Om03}
   \end{center}
\end{figure}

After computing the power spectrum from eqn. \ref{eqn_Cl_cluster} for 
hundreds of models where we change $\sigma _8$ and $\Omega$, 
we can compare the predicted power spectrum with the 
observed RASS power spectrum in the R6 band and exclude all the models for which 
we have more predicted power than the observed one. 
We show the result in figure \ref{fig_Om_S8}. The dotted line is a numerical 
fit to the upper limit. 
\begin{equation}
\sigma_8 = 0.5\Omega_m^{-0.38}
\label{eqn_S8Om}
\end{equation}
Our limit excludes regions in the $\sigma _8 - \Omega _m$ plane which are within the 
confidence region of many other methods.  
In figure \ref{fig_ROSAT_Cl_S808_Om03} we compare the RASS power spectrum with 
three models above, on,  and below the upper limit curve ($\sigma_8 = 0.75, 0.8, 0.85$, 
for $\Omega = 0.3$). 
The range of $l$'s which exclude the models above the upper limit are 
up to $l \approx 200$ which corresponds to angular scales 
$\theta \approx 1^{\circ}$. At these scales only the nearby clusters 
(like the Abell clusters) contribute to the power spectrum (see figure \ref{fig_Cl_z} and 
\ref{fig_Cl_mass}). Hence, with this data we are not very sensitive 
to the population of clusters at moderate and high-redshift. 
This fact is in agreement with the measured cross-correlation between the angular position 
of Abell clusters and the intensity of the X-ray background (Soltan et al. 1996). 
In that work, the authors found that the cross-correlation extends up to several degrees  
({\it i.e.} the cluster--X-ray background cross-correlation spreads over the range 
$l < 100$). \\
The RASS data still contains some contribution of non-removed point sources 
which could be contaminating the power spectrum at sub-degree scales. 
This fact suggests that a study carried out on a smaller 
region of the sky with a more careful point source removal could render better constraints 
on the cosmological model. 
This opens an exciting possibility for future work carried out with $Chandra$ and/or 
especially with XMM-Newton data. XMM-Newton has a higher sensitivity
for extended sources and wider field of view 
than $Chandra$. This makes XMM an ideal instrument for wide field X-rays surveys (like the 
XMM-LSS, Refregier et al. 2002a). 
However, the analysis of the arcmin-subarcmin scales is beyond the scope of this paper 
and will be discussed in a subsequent paper.\\

The constraints derived in this section are robust since they were obtained under a  
minimum number of assumptions. \\
However, variations in these assumptions ($L_x-T$, core radius) can have 
an important effect on the resulting power spectrum.
It is important to study how the upper limits obtained in this 
section can be affected by varying our assumptions ($L_x-T$ relation and 
core radius, $r_c$). Although the model we assumed for $L_x-T$ and core radius 
was consistent with several optical and X-ray cluster data sets 
(as shown in Diego et al. 2001), there are some uncertainties in the 
values of the parameters assumed in the model 
($L_o$, $\alpha$, $\psi$ and $r_c$). A different set of values for these 
parameters will render a different upper limit on the  $\sigma _8 - \Omega _m$ 
plane. Our results can be therefore affected by systematic errors due to the wrong  
election of the parameters of the model. In the next section we will 
attempt to see how much does our upper limit change when we make different 
assumptions in our model.\\
\section{Systematic effects}\label{sect_systematic}
\begin{figure}
   \begin{center}
   \epsfysize=6.cm 
   \begin{minipage}{\epsfysize}\epsffile{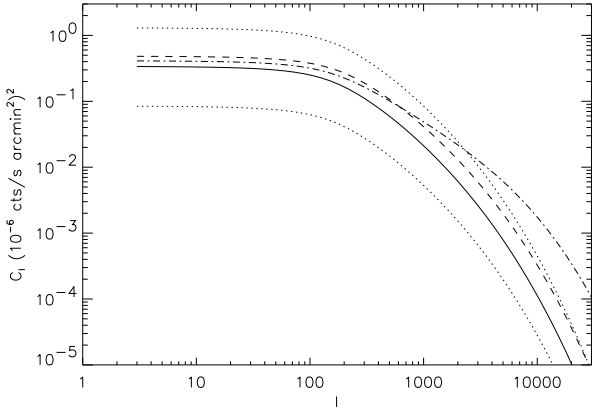}
   \end{minipage}
   \caption{
           Systematic effects. Effect of changing $L_o$, $\alpha$ and $\psi$ in the 
           $L_x - T$ relation (eqn. \ref{eqn_ST}) for a $\sigma _8 = 0.8$, 
           $\Omega _m = 0.3$ model. 
           The thick solid line is the reference model. 
          ($L_o = 1.12\times 10^{42}  h^{-2} erg/s$, $\alpha = 3.2$, $\psi = 1$). 
           The dashed line correspond to the same value of $L_o$ but $\alpha = 2.7$. 
           The dotted lines are for $L_o = 2.25\times 10^{42}  h^{-2} erg/s $ 
           with $\alpha = 3.2$ 
           (top) and $L_o = 0.56\times 10^{42}  h^{-2} erg/s $ with $\alpha = 3.2$. 
           In the dot-dashed line we fix $L_o = 1.12\times  h^{-2} 10^{42} erg/s$, and 
           $\alpha = 3.2$ and we change $\psi$ in four units, $\psi=5$. 
           }
   \label{fig_SystemEffect_Lx_alpha}
   \end{center}
\end{figure}

The main source of possible systematic error is the normalisation of 
the assumed $L_x-T$ relation, $L_o$ (eqn. \ref{eqn_ST}). 
There is still an important uncertainty in this relation and even a
break could be present at low temperatures. This break in the scaling 
has been interpreted as due to preheating phenomena (e.g. Voit \& Bryan 2001, 
Babul et al. 2002).
Since the power spectrum of the clusters is proportional to the square of 
the total signal (eqn. \ref{eqn_po}).  The power spectrum 
will be therefore very sensitive to the normalisation of the $L_x-T$ relation 
($L_o$ in eqn. \ref{eqn_ST}). This is illustrated in figure 
\ref{fig_SystemEffect_Lx_alpha} where we change $L_o$ by a factor of two and 
the power spectrum changes by a factor of four (dotted line) as expected. 
The dependence of the power spectrum on the scaling exponent, $\alpha$, is not 
as dramatic as in the case of $L_o$. When we change $\alpha$ between  $\alpha=3.2$ 
and $\alpha=2.7$ the power spectrum changes from the solid line to the dashed line.
Finally, the factor controlling the redshift dependence, $\psi$, is not very relevant 
up to $l \approx 100$ but at smaller scales ($l > 100$) it can become important since as 
we discussed in section \ref{sect_theor_Cl}, 
at the smallest scales the power spectrum is dominated by the high redshift population.
The previous plot demonstrates that one must be careful in choosing the 
$L_x-T$ relation since the results can change significantly. \\

The second assumption on our model is about the 2D profile. In this case, our only free 
parameter is the scaling of the core radius, $r_c$. 
We will study the effect of changing the core radius by changing the ratio 
$r_o/p$ in equation \ref{eqn_rc}. 
\begin{figure}
   \begin{center}
   \epsfysize=6.cm 
   \begin{minipage}{\epsfysize}\epsffile{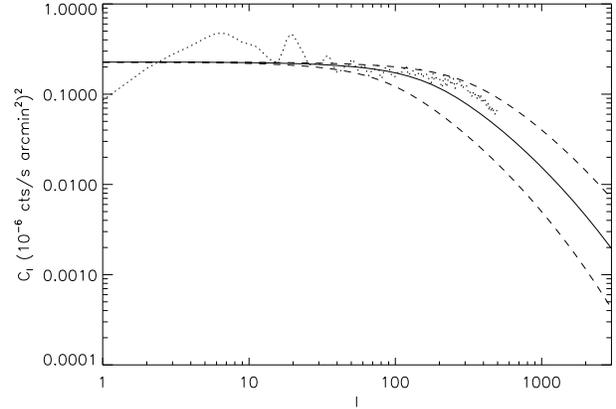}
   \end{minipage}
   \caption{
            Systematic effects. Effect of changing the parameter $p$ in equation 
            \ref{eqn_rc} while fixing $r_o = 130 h^{-1}$ Mpc. The solid line is the 
            reference model with $p=10$ ($\sigma _8 = 0.77$, $\Omega _m = 0.3$). 
            The two dashed lines are for $p=5$ (bottom) 
            $p=20$ (top). As expected, clusters which are more extended ($p= 5$) 
            have less power at smaller scales but do not change the power in the 
            large scales (see text). 
            On the other hand, clusters which are more clumped than the $p=10$ models 
            help to increase the power at smaller scales (without affecting the 
            large scales) and they can produce a good match to the shape of the 
            observed X-ray power spectrum (dotted line) at high multipoles 
            ($\theta < 30$ arcmin). This fact suggests that the clumpiness parameter, 
            $p$, could be in the range $10 < p < 20$. 
           }
   \label{fig_SystemEffect_p}
   \end{center}
\end{figure}
The effect of changing the scaling of core radius is only important at small scales 
(large $l$). This is not surprising since at low $l$'s the value of power spectrum is 
driven only by its normalisation ($p_o$ in equation \ref{eqn_pl}). When we change the 
parameter $p$ (ratio between the virial and core radius), we observe that clusters 
with larger $p$ (i.e smaller core radius when $r_o$ is fixed) have more power 
at smaller scales (figure \ref{fig_SystemEffect_p}). This point is interesting since it 
shows how our constraints on $\sigma _8 - \Omega _m$ are not very sensitive to the 
cluster's profile. The $L_x-T$ relation will be the main source of uncertainty.\\

\begin{figure}
   \begin{center}
   \epsfysize=8.cm 
   \begin{minipage}{\epsfysize}\epsffile{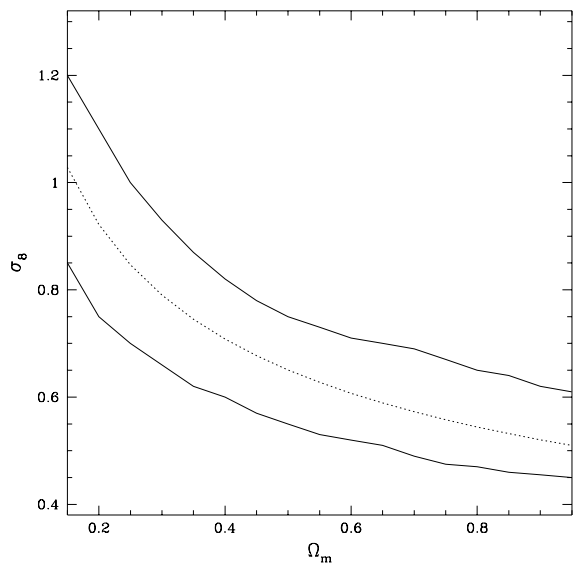}\end{minipage}
   \caption{
            Upper limits constraint in the $\sigma _8 - \Omega _m$ plane for the 
            optimistic (bottom solid line) and the pessimistic (top solid line) 
            models. The dotted line is the numerical fit obtained in section 
            \ref{sect_constraints} ($\sigma_8 = 0.5\Omega_m^{-0.38}$). 
           }
   \label{fig_Om_S8_Systematic}
   \end{center}
\end{figure}

From the previous results we can conclude that due to our uncertainty in 
the $L_x-T$ relation and the cluster density profile (parametrised as a function of $r_c$) 
we have an uncertainty in the theoretical power spectrum of clusters of $\approx 1$ order 
of magnitude. We will incorporate this uncertainty in our upper limit in the 
$\sigma _8 - \Omega _m$ plane by rescaling our theoretical model by a factor of 
5 in both directions (multiplication and division). 
By doing this we will have an {\it optimistic} and a {\it pessimistic} 
model. The optimistic model will contain 5 times more power than the model assumed 
in section \ref{sect_constraints} while the pessimistic one will contain 5 times less power.
The difference in power between the optimistic and the pessimistic is then a factor 25.
The result is shown in figure \ref{fig_Om_S8_Systematic}.
Even in the pessimistic case, the constraints are still important (models 
like $\sigma _8 = 1$, $\Omega _m = 0.3$ can be excluded). As an example, the 
pessimistic case exclude more or less half of the confidence region obtained from the 
REFLEX cluster power spectrum (see figure 4 in Schuecker et al. 2002). \\
It is more interesting to see what happens in the optimistic case. Here, models with 
$\sigma _8 = 0.8$, $\Omega _m = 0.3$ would be excluded. This particular model 
is actually one of the best in fitting several recent data sets (CMB, 2dF, etc). 
This means that, if the underlying cosmological model is in fact $\sigma _8 = 0.8$, 
$\Omega _m = 0.3$, our optimistic case is {\it too} optimistic!
This opens the possibility of using the diffuse X-ray background as a way to 
study cluster physics rather than constraining the cosmological model. 
In the hypothetical scenario where the cosmological model is known, one 
can use the X-ray power spectrum to constrain the physical parameters of the 
intra-cluster plasma. If, for the {\it known} cosmology, one observes that the 
theoretical cluster power spectrum exceeds the observed one, this means 
that the assumptions made on the modelling of the cluster power spectrum were 
wrong (too optimistic case). 
One should then reintroduce changes in the model in order 
not to exceed the observed X-ray power spectrum (pre-heating, electron density 
profile, evolution of the typical luminosity with redshift, baryon fraction, etc)
\section{Beyond the cosmological model}\label{sect_beyond}
The main difference between the constraints obtained in this work and those obtained by other 
methods (based on galaxy clusters) is that we can have a good control on the systematic 
errors. 
Our results are not affected by a changing selection function in the
sky, or extrapolation of 
the observed central cluster 2D profile beyond the noise level.

There are only two key assumptions in our model, the $L_x-T$ relation and the cluster density 
profile. At the scales relevant in this work ($l < 100$) the uncertainty in the 
cluster density profile is not very important. On the contrary, the main uncertainty comes 
from our partial knowledge of the  $L_x-T$ relation. Most of the {\it cosmological} 
methods based on galaxy clusters data obtain the best fitting cosmological model 
assuming that the physics of clusters are well known. If the uncertainties in the 
cluster physics are not included, then a biased estimate of the best cosmological model 
may emerge as a result. 
We are approaching an era where the best cosmological model is quickly 
converging to a set of {\it preferred values}. One can then think the other way around, 
assuming we know the cosmological model, what can we infer about the physics in clusters ?
\subsection{Preheating and the $L_x-T$ relation}
\begin{figure}
   \begin{center}
   \epsfysize=6.cm 
   \begin{minipage}{\epsfysize}\epsffile{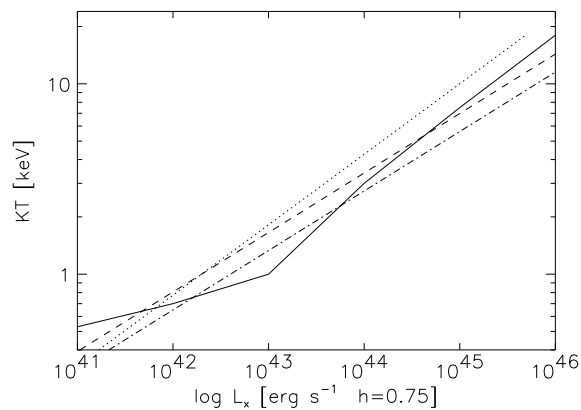}\end{minipage}
   \caption{$L_x-T$ relation compared with the pre-heated model (solid line) in 
            Babul et al. (2002). The dashed line represents the main model used 
            in our analysis (figure \ref{fig_Om_S8}). That model was the best fitting 
            model found in Diego et al. (2001) ($L_o = 1.12 \times 10^{42}  h^{-2} erg/s$, 
            and $\alpha = 3.2$, see eqn. \ref{eqn_ST}) after fitting the cluster mass function, 
            temperature function, X-ray luminosity function and X-ray flux function. 
            Also shown are two of the models used in figure \ref{fig_SystemEffect_Lx_alpha}. 
            Dotted line is for $L_o = 1.12 \times 10^{42}  h^{-2} erg/s$, and $\alpha = 2.7$, and 
            dot-dashed line is for $L_o = 2.25 \times 10^{42} h^{-2}  erg/s$, and $\alpha = 3.2$. 
            These two models also produced a reasonable good fit in Diego et al. (2001). 
            In he above plot we have assumed $h=0.75$ in $L_o$ which was the value adopted in 
            Babul et al. (2002). All the models represented in this plot give a reasonable fit 
            to the observed $L_x-T$ relation (see figure 5 in Babul et al. 2002). 
           }
   \label{fig_LT}
   \end{center}
\end{figure}

In figure \ref{fig_LT} we present some of the estimates of the $L_x-T$ relation used 
in this work compared to a model including pre-heating (solid line, Babul et al. 2002). 
All the models represented in the plot make a good fit to the most recent estimations of 
the $L_x-T$ relation (see e.g, Markevitch 1998, Allen \& Fabian 1998, Helsdon \& Ponman 2000 ). 
The pre-heating model corresponds to an entropy floor of $kTn_e^{-2/3}
\approx 427\, {\rm keV}\, {\rm cm}^2$. 
This pre-heated model gives an excellent match to the observations (see Babul et al. 2002).
The model used in section \ref{sect_constraints} corresponds to the dashed line. 
The dotted line and dot-dashed line where used in section \ref{sect_systematic} 
to illustrate the change in the cluster power spectrum when the parameters 
$L_o$ and $\alpha$ where changed.
The main difference between the pre-heated model and the single scaling relation is that 
at temperatures $T \approx 1$ keV the pre-heated models predict a larger luminosity than the 
standard single scaling relations. Clusters with $T$ of a few keV correspond to 
intermediate masses. We have seen that this range of masses dominate the power spectrum so 
we should expect some kind of dependence on the power spectrum with the amount of entropy 
in the pre-heated model. As shown in Babul et al (2002), lower entropy floors will produce 
a larger shift of the relation to higher X-ray luminosities (lower $\alpha$) while higher 
entropy floors will result in a steeper $L_x-T$ relation (higher $\alpha$).  
At $kT$ below 1 keV the situation inverts and the pre-heated model predicts clusters to be 
less luminous than in the single scaling case.
Previous studies based on the intensity of the soft X-ray background suggest 
that there must be some level of pre-heating on galaxy clusters, for
example from AGN heating of the intracluster gas. Otherwise,  
their contribution to the soft to X-ray background exceeds the derived upper limit 
(see e.g. Wu et al. 2001). 
\begin{figure}
   \begin{center}
   \epsfysize=6.cm 
   \begin{minipage}{\epsfysize}\epsffile{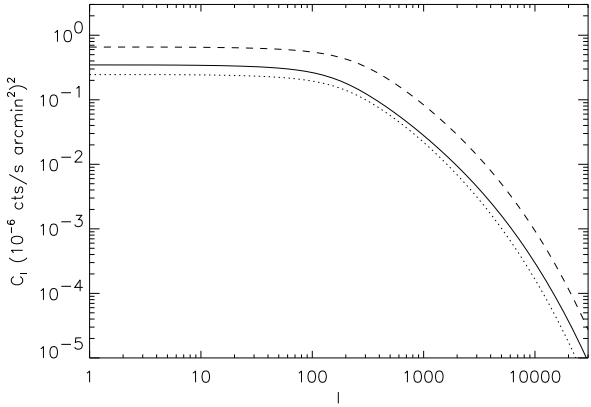}\end{minipage}
   \caption{
            Change in power when pre-heating is considered. The solid line is the reference 
            model ($\sigma _8 = 0.8$, $\Omega _m=0.3$, $L_o = 1.12 \times 10^{42}  h^{-2} erg/s$, 
            $\alpha = 3.2$ and $\psi = 0.0$. In the reference model the temperature of the 
            cluster is computed from the $T-M$ relation obtained in Diego et al. (2001) 
            ($T(keV) = 9.48 M_{15}^{0.75}(1 + z)$). The dashed line shows the power spectrum 
             when the X-ray luminosity is computed from the pre-heating model 
            (Babul et al. 2002) shown in figure \ref{fig_LT} and the temperature is obtained 
             from a numerical fit to the $T-M$ relation (Nevalainen et al. 2000), 
            $T(keV) \approx 8.0 M_{15}^{0.56}$. 
            For reference (dotted curve), we also show the cluster X-ray power 
            spectrum when we use the Nevalainen et al. (2000) $T-M$ relation but the 
            non-preheated model is used to get the X-ray luminosity (i.e. 
            $L_o = 1.12 \times 10^{42}  h^{-2} erg/s$ and $\alpha = 3.2$ for all $T$)
            }
   \label{fig_preheating}
   \end{center}
\end{figure}
In figure \ref{fig_preheating} we show how the preheating can change the power spectrum 
of clusters. The solid line is the non-preheated model used in section \ref{sect_constraints} 
but with non-evolving $L_x-T$ ($\psi = 0$) (dashed line in figure \ref{fig_LT}). When we use 
the pre-heated model of Babul et al. (2002) (solid line in figure \ref{fig_LT}) we see an 
increase in the power spectrum by a factor of $\approx 2$ (dashed line). 
In the pre-heated model case we also have computed the temperature of 
the cluster (from its mass) using a numerical fit to the observed $T-M$ relation 
(Nevalainen et al. 2000). That fit differs from the best-fitting $T-M$ function found in 
Diego et al. (2001). 
The Nevalainen's $T-M$ relation is in perfect agreement with the 
latest estimate  of the $T-M$ scaling (Shimizu et al. 2002) where the authors  
find $kT({\rm keV}) = 7.7 M_{15}^{0.55}$ ($M_{15}$ is expressed in $10^{15} h^{-1} M_{\odot}$). 
It is also in agreement with previous results which suggest that the $T-M$ scaling relation 
differ from the expected self-similar behaviour (Mohr et al. 1999).
When we use Nevalainen's $T-M$ (but keeping the rest of parameters of the 
model corresponding to the solid line in figure \ref{fig_preheating}) the new power spectrum changes 
from the solid line to the dotted line. \\

If we assume that the pre-heating model used above is a good description of the 
$L_x -T$ relation, we can extract some surprising conclusions.
Our upper limits exclude many of the previous constraints on this parameter 
(see Refregier et al. 2002b for a recent compilation of constraints on 
$\sigma _8$). From the results compiled in Refregier et al. (2002b), 
only the recent 2dF$+$CMB (Lahav et al. 2001) and 
galaxy clusters (Seljak 2002) constraints are well inside our upper limits. This can have two 
interpretations. The first interpretation is that those constraints derived using  
other methods which are in contradiction with the RASS upper limit on
$\sigma _8$ need to be  revised with respect to 
the assumptions made in the appropriate analyses. 
The second interpretation is that we are being 
too optimistic in our assumptions. The assumed $L_x-T$ relation is not a good description of the 
underlying $L_x-T$ relation in galaxy clusters and we may need to {\it inject} more entropy in 
the clusters in order to make them less luminous (or any other mechanism which decreases the 
X-ray luminosity). This second interpretation is interesting because 
it shows us how we can study cluster physics if we know something
about the cosmological model.
\begin{figure}
   \begin{center}
   \epsfysize=8.cm 
   \begin{minipage}{\epsfysize}\epsffile{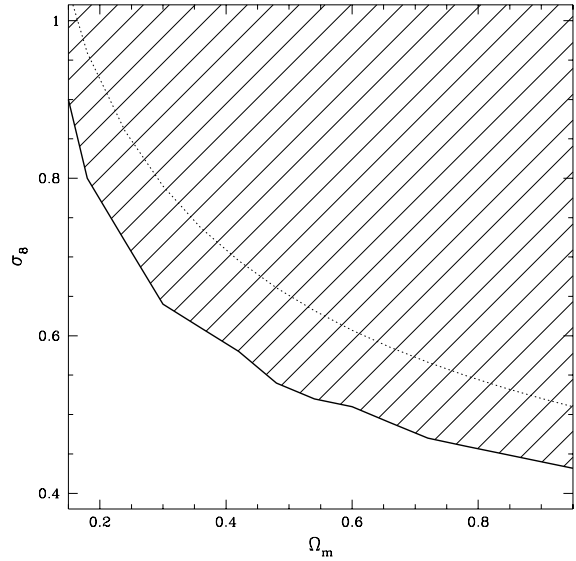}\end{minipage}
   \caption{
            $\sigma _8 -\Omega _m$ constraints assuming the pre-heating model of Babul et al. 
            (2002) and the $T-M$ relation of Nevalainen et al. (2000), (thick solid line). 
            This upper limit would exclude models like $\sigma _8 = 0.8$, $\Omega _m = 0.3$.
            For reference we also show the result obtained in section \ref{sect_constraints} 
            (dotted line). 
            }
   \label{fig_S8_Om_preheating}
   \end{center}
\end{figure}
If we assume that the underlying cosmological model can be described by the parameters 
$\sigma _8 = 0.8$ and $\Omega _m = 0.3$ (in a flat $\Lambda$CDM universe with $\Gamma = 0.2$) 
then we can ask whether or not the assumed physics ($L_x-T$ relation) is in agreement with the 
observations. We can test this by imposing that the theoretical power spectrum must be below 
the observed one. 
\subsection{$L_x-T$ and $T-M$ from $\sigma _8 - \Omega _m$}
\begin{figure}
   \begin{center}
   \epsfysize=8.cm 
   \begin{minipage}{\epsfysize}\epsffile{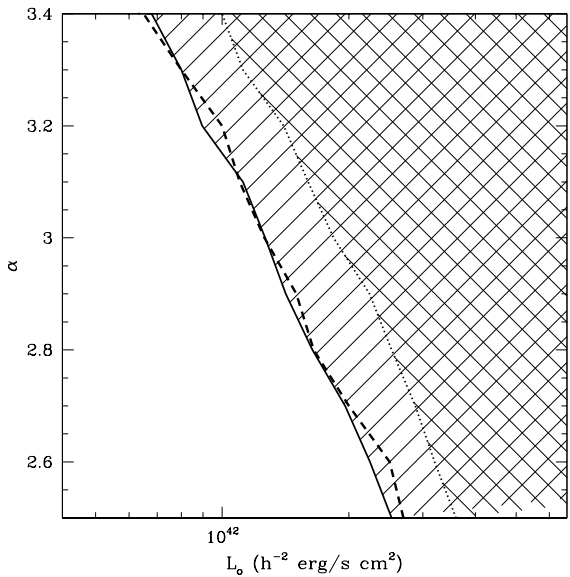}
   \end{minipage}
   \caption{
            Upper limits on the value of $L_o$ as a function of $\alpha$ for a 
            cosmological model with $\sigma _8 = 0.8$ and $\Omega _m = 0.3$. 
            The solid line mark the upper limits when we assume 
            $T = 8.5 M_{15}^{0.54}$ (Nevalainen et al. 2000). 
            If the assumed $T-M$ relation is $T = 8.0 M_{15}^{2/3}$ (Evrard et al. 1996) 
            the upper limits on $L_o$ change (dotted line). 
            We also show the case when a pre-heating like model with two different exponents 
            is assumed (dashed line), $L_x = L_o *T^{\alpha}$ for $T > 1$ keV and 
            $L_x = L_o *T^{\beta}$ for $T < 1$ keV. We assumed $\beta = 2\alpha$ and 
            the Nevalainen et al. (2000) $T-M$ relation. This shows how the constraints are 
            not very sensitive to clusters with $T < 1$ keV.
            }
   \label{fig_constraint_LT}
   \end{center}
\end{figure}

Although there is still a very lively  debate about the value of $\sigma _8$, there is a 
general consensus in the allowed range for $\sigma _8$, $\sigma _8 \in (0.7-1.1)$ 
for $\Omega _m = 0.3$ with a most favoured value of $\sigma _8 \approx 0.8$. 
The fact that the preheating model considered above is excluding this most favoured 
value can have a simple interpretation, which is that 
our assumptions about the luminosities of the clusters are too optimistic. 
The clusters must be less luminous than  we assumed in order to allow the upper 
limit to stay above the preferred value of $\sigma _8$. 
If we parametrise the luminosity as $L_x = L_o T^{\alpha}$, then we can constrain $L_o$ as 
a function of $\alpha$ by just requiring that the cluster power spectrum stays below the 
measured power spectrum (see figure \ref{fig_constraint_LT}).
Although we have not discussed much about the $T-M$ relation, it will also play a role 
in the definition of the upper limit. To get the temperature in the $L_x - T$ relation we 
need to assume some form for $T-M$ relation. Recent observations (Nevalainen et al. 2000), 
seem to indicate that this relation differs slightly from the self similar scaling relation. 
However, as in the case of the $L_x-T$, there is some scattering on
the $T-M$ relation. 
We have included this uncertainty into our calculations by using two different 
$T-M$ relations. In the first case we use the fit to the observed $T$ and $M$ obtained 
by Nevalainen et al. (2000), $T = 8.5 M_{15}^{0.54}$ keV and in the second case we use 
a self-similar form (Evrard et al. 1996), $T = 8.0 M_{15}^{2/3}$ keV. 
The results on $L_o$ and $\alpha$ are shown in figure \ref{fig_constraint_LT}.\\
    
\begin{figure}
   \begin{center}
   \epsfysize=6.cm 
   \begin{minipage}{\epsfysize}\epsffile{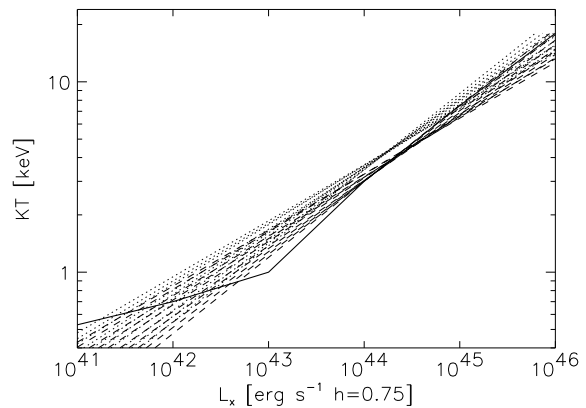}
   \end{minipage}
   \caption{
            $L_x-T$ relations for the models lying in the two upper limits of 
            figure \ref{fig_constraint_LT}. The dotted line corresponds to the 
            models in the solid line of figure \ref{fig_constraint_LT} while the 
            dashed lines correspond to the models in the dotted line curve 
            in figure \ref{fig_constraint_LT}. The solid line shows the prehetaing 
            model of Babul et al. (2002). 
            }
   \label{fig_LT_plusupperlimits}
   \end{center}
\end{figure}

When we compare the models on the two upper limit curves with the observed $L_x-T$ 
relation (figure \ref{fig_LT_plusupperlimits}),  
we find that the models on the upper limits are marginally consistent with the observations. 
The model with the self-similar $T-M$ relation produces a constrain which agrees better 
with the observations.
\section{Discussion}\label{sect_discussion}
We have found that in order to explain most of the constraints obtained in $\sigma _8$ 
and $\Omega _m$, we need a rather steep $L_x-T$ relation. We have changed these 
relations within the observed limits (and even more, in the case of the parameter 
$\alpha$) and we found that the $\sigma _8 = 0.8$  $\Omega _m = 0.3$ model can be accommodated 
only in the most extreme cases. Using the pre-heating model of Babul et al. (2002), our 
upper limit excludes most of the current constraints in $\sigma _8$ for $\Omega _m = 0.3$. \\
If $\sigma _8 > 0.8$  for $\Omega _m = 0.3$, we may say that there is an inconsistency 
between the above preheating model and the upper limit imposed by the RASS diffuse 
background. 

As noted by Voit et al. (2002), the observed $L_x-T$ 
relation changes when the luminosities are corrected by cooling flow effects. 
Our results suggest that cooling flows can be important (specially at low redshift which 
is the range dominating the power spectrum in this scale). 
A $L_x-T$ which is obtained without correcting for cooling flows will be affected by a 
larger scatter due to the presence of the clusters with cooling flows 
(they will increase the average luminosity of clusters and therefore the value of $L_o$). 
If cooling flow clusters are not representative of the average population of clusters, 
they should be excluded from any fit to the $L_x-T$ relation or corrected their luminosities 
(and average temperatures) by the effect of the cooling flow in order to get a representative 
$L_x-T$ relation (see Fabian et al. 1994 for a discussion on the effect of cooling flows 
in the $L_x-T$ relation).\\

\begin{figure}
   \begin{center}
   \epsfysize=8.cm 
   \begin{minipage}{\epsfysize}\epsffile{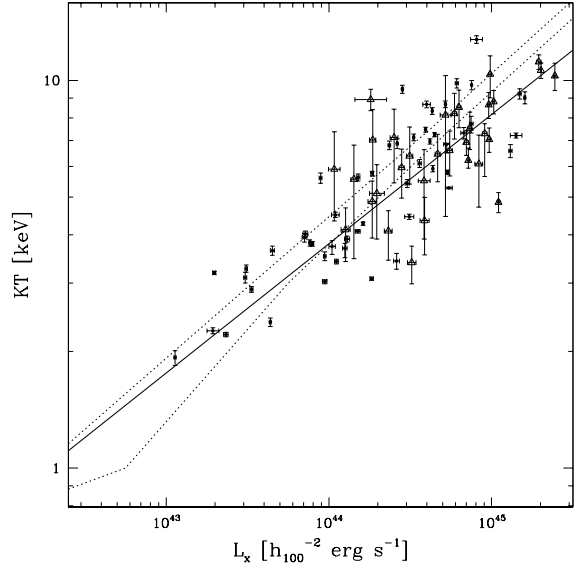}
   \end{minipage}
   \caption{
            Scaling model for the curve in the upper limit $L_o= 2.3 10^{42} h^{-2} erg/s$ and 
            $\alpha = 3$ (solid line). The model is compared with a recent compilation of 
            temperatures and luminosities (Novicki et al. 2002) at low (dots) and intermediate 
            redshift (open triangles). The two dotted lines show the preheating models of 
            Voit \& Bryan (2001) (top) and Babul et al. (2002) (bottom). 
            Note the large scattering in the data.
            }
   \label{fig_LT_plusupperlimitsB}
   \end{center}
\end{figure}

In figure \ref{fig_LT_plusupperlimitsB} we present one of the {\it extreme} cases  
for which the model $\sigma _8 = 0.8$, $\Omega _m = 0.3$ is just below the upper 
limit (solid line). This model corresponds to the values  $L_o= 2.3
\times 
10^{42} h^{-2}\, {\rm erg}\, {\rm s}^{-1}$ 
and $\alpha = 3$ and is in the dotted curve of figure \ref{fig_constraint_LT} 
(i.e. when we assume the self-similar relation $T = 8.0 M_{15}^{2/3}$ keV). If we compare this 
extreme case with real observations of the $L_x-T$ relation, we see that this model 
still produces a good fit to the data (but we need to assume that $T \propto M_{15}^{2/3}$). 
All the models lying on the right hand side of the solid line overpredict the observed RASS power 
spectrum for $\sigma _8 = 0.8$, $\Omega _m = 0.3$. 
The pre-heating model of Babul et al. (2002) (bottom dotted line) is marginally consistent 
with the upper limit above 4 keV but exceeds the upper limit below that temperature. 
On the other hand, the pre-heated model of Voit \& Bryan (2001) is still 
compatible with the upper limit but is well within the limits. 
There is little room for additional changes in the $L_x-T$ and $T-M$ relation without 
coming into serious contradictions with the observed relations. As a consequence, 
our results suggest that it is very difficult to accommodate high $\sigma _8$ models 
($\sigma _8 \geq 0.8$) within our upper limit. Recently, a combined analysis of 2dF plus CMB 
data has produced new constraints on $\sigma _8$ which suggest that $\sigma _8 \approx 0.7$ 
for $\Omega _m h = 0.21$ (no tensor model) and $\sigma _8 \approx 0.6$ for 
$\Omega _m h = 0.16$ (tensor model) (see Efstathiou et al. 2002). 
Other recent works also suggest a low value for $\sigma _8$ (e.g Allen et al. 2002, 
Seljak 2002, Bahcall et al. 2002).  
These values of $\sigma _8$ are in very good agreement with our upper limits. 
If the value of $\sigma _8$ is significantly larger than 0.8, (for $\Omega _m = 0.3$) 
then, we can conclude that there is a serious contradiction between the X-ray data and the 
model. 
If we adopt the {\it realistic} position that the inconsistency must be in the model, 
then, we should revise our assumptions in order to resolve this puzzling situation.\\
The only assumption we have not checked in our model is that about the mass function. 
As mentioned in the introduction, the modelling of the cluster power spectrum involves 
the whole population of clusters. This fact makes the cluster power
spectrum a very sensitive probe of the cluster mass function. \\
This opens the exciting possibility of (assuming the cosmological model is known) 
constraining the cluster mass function once we know accurately the $L_x-T$ and $T-M$ 
relations and a good estimation of the X-ray background power spectrum is available. \\

There is an important point which has not been explored in this paper 
but that we plan to study in a subsequent work. It is the scattering of the 
$L_x-T$ relation. Real clusters will show an intrinsic scattering in this relation 
(clusters with the same temperature may have different luminosities around the 
{\it average} $L_x$). 
Since the power spectrum of clusters is proportional to the square of the 
multipole decomposition (equations. \ref{eqn_Cl_cluster} and \ref{eqn_po}), 
those clusters with the same $T$ but $L_x$ larger than the mean 
will contribute more to the power spectrum than the clusters with $L_x$ lower than the mean 
(but the same $T$). That is, the {\it average} effect of the population 
of clusters with the same $T$ (but different $L_x$) is not the same as the effect 
of the population of clusters with the same $T$ and {\it average} luminosity $L_x$ 
(due to the quadratic dependence of $C_l$). This fact may change the results presented 
in this work and need to be studied in more detail. 
\section{Conclusion}\label{sect_conclusion}
\begin{figure}
   \begin{center}
   \epsfysize=6.cm 
   \caption{
             Upper limit on $\sigma _8 - \Omega _m$ compared with a recent 
             constraint in these parameters from CMB (Melchiorri et al. 2002). 
             The X-ray backgound upper limit excludes half of the CMB constraints.
            }
   \label{fig_CMB}
   \end{center}
\end{figure}
We have shown that the use of the measured X-ray power spectrum as an upper limit produces
strong constraints on the cosmological model.
The constraints have the advantage of being very robust since they were obtained 
under a minimum number of assumptions. The uncertainty in the assumptions can be easily 
incorporated into the model to produce optimistic and pessimistic constraints. 
Approaches like the one presented in this work are needed to complement the constraints 
obtained using other methods (see figure \ref{fig_CMB}).
To obtain the upper limits on $\sigma _8 - \Omega _m$ presented in this work we do not 
even need to detect the clusters. The global contribution of all the clusters (above or 
below the $3\sigma$ level) contributes to the power spectrum although we have seen 
that the cluster population with intermediate masses and $z < 0.2$ dominates the power spectrum 
in the range of scales considered in this work.\\
In this work, we have not considered the clustering contribution (two-halo). This contribution 
may be of the order of $20\%-30\%$ at the scales relevant for this work 
(Lahav et al. 1997, Komatsu \& Kitayama 1999). 
The upper limits presented here will be even more significant 
if this contribution is incorporated into the analyses. \\
Due to the point source contamination and large pixel size, 
the constraints on $\sigma _8 - \Omega _m$ arise in the range $l < 100-200$. 
However, this fact has an advantage. At low $l$'s, the cluster power 
spectrum is independent of the electron density profile (or the internal structure 
of the cluster) and only the total luminosity of the clusters matters for modelling the power 
spectrum. This reduces dramatically the number of assumptions made in the model. 
One may use this fact to make studies of the physics of the intracluster plasma 
(like pre-heating) which could constrain the $L_x-T$ relation.\\
We found that, in order to accommodate cosmological models like $\sigma _8 = 0.8$  
$\Omega _m = 0.3$ we need to {\it tune} the $T-M$ and the $L_x-M$ relation to the limits 
of the observational constraints. This fact suggest that lower $\sigma _8$ models 
are favoured by the upper limit constraint set by the X-ray background.

The approach presented in this work can be extremely useful with
higher resolution and better signal to noise data 
($Chandra$, and particularly XMM-Newton). 
Although the data in the $ROSAT$ R6 band has the brightest point sources 
removed,  the power spectrum still contains some contribution coming from 
the non-removed point sources and their clustering. 
Future data from $Chandra$ and specially XMM-Newton will allow a better point source 
subtraction. The constraints obtained from these new data will be much more significant 
since one can probe smaller scales with a significantly smaller contamination from 
point sources. 
In the small scale range, the cluster power spectrum becomes even more sensitive to 
the physics of the plasma (angular extension of the emission and evolution of the $L_x-T$ 
relation). In particular, we have seen that the shape of the power spectrum at small scales 
(large multipoles) is better described by density profiles having high clumpiness factors  
($p \approx 10-20$). 
A low-contaminated point source power spectrum could lead to dramatic 
constraints on the $L_x-T$ relation. These constraints can have serious implications for the 
physics of the intracluster plasma (pre-heating, cooling flows). 
These facts make worth extending the analyses presented here to smaller scales with a high 
sensitivity where the point source removal can be performed to very low fluxes.
\section{Acknowledgements}
This research has been supported by a Marie Curie Fellowship 
of the European Community programme {\it Improving the Human Research 
Potential and Socio-Economic knowledge} under 
contract number HPMF-CT-2000-00967. XB acknowledges financial support
from the Spanish Ministry of Science and Technology under project AYA2000-1690.
The authors would like to thank Wayne Hu, Greg Bryan and Francisco Carrera 
for fruitful discussions. We also thank A. Melchiorri for providing 
the plot with the confidence levels derived from CMB.



\bsp
\label{lastpage}
\end{document}